# N-Body Simulation of the Formation of the Earth-Moon System from a Single Giant Impact

Justin C. Eiland[1], Travis C. Salzillo[2], Brett H. Hokr[3], Justin L. Highland[3] & Bryant M. Wyatt[2]

**Summary**
The giant impact hypothesis is the dominant theory of how the Earth-Moon system was formed. Models have been created that can produce a disk of debris with the proper mass and composition to create our Moon. Models have also been created which start with a disk of debris that eventually coalesces into a Moon. To date, no model has been created that produces a stable Earth-Moon system in a single simulation. Here we combine two recently published ideas in this field, along with a new gravity-centered model, and generate such a simulation. In addition, we show how the method can produce a heterogeneous, iron-deficient Moon made of mantle material from both colliding bodies, and a resultant Earth whose equatorial plane is significantly tilted off the ecliptic plane. The accuracy of the simulation adds credence to the theory that our Moon was born from the violent union of two heavenly bodies.

**Expansion of summary**
Prior to landing on the Moon, three theories dominated the origins of the Moon debates[1-3]. First, could the Moon be a twin planet to Earth formed out of the same cloud of gas and dust? If so, their overall compositions would be very similar[1-3]. However, the Moon has a relatively small iron core when compared to Earth, so this is unlikely. Second, is the Moon a captured rocky planet? If this were true, the Moon's composition should be unlike Earth's. The Moon has striking similarities to Earth in several isotopes[4-11], therefore, this is also unlikely. Third, could a young, fast-spinning Earth have spun off a large section of its mantle to have formed the Moon? This would explain the Moon's iron deficiency and similarities to Earth's mantle. However, the Moon's absence of volatiles suggests that it originated in a hotter environment than Earth[1-3], leaving this fission theory flawed as well.

    The origin of Earth's Moon continues to be a heavily debated topic in astrophysics today. How could the Moon have many similar isotopes to Earth, yet have a shortage of volatiles and lack a large amount of iron[3]? In 1975, Hartmann and Davis and, independently in 1976, Cameron and Ward proposed the giant impact hypothesis which seems to answer this question[1,12-15]. The hypothesis states that the Earth-Moon system was formed by the collision of the early Earth and a Mars-sized planet called Theia. Their heavy iron cores coalesced, and a large amount of crust and mantle material were violently ejected, leaving the system in a state of high angular momentum. The ejected material is thought to have created a circumplanetary disk around Earth which later accreted into the Moon[12-16]. Computer simulations added credibility to this theory by numerically demonstrating that moon formation from a giant impact was indeed feasible[1,15,17]. Most numerical simulations of this theory focus on creating a disk of debris with the proper mass and composition which could lead to the formation of the Moon[5,6,18,19]. Other simulations start with a disk of debris and demonstrate how this disk can coalesce into our Moon[17,20]. However, no model has produced a stable Earth-Moon system from a single simulation. Recently, the focus of simulations has been to explain the similarity in composition between Earth and the Moon. Robin Canup of the Southwest Research Institute has been studying collisions with two Earth-sized impactors and has shown how this would eject more material from Earth's mantle into orbit[6]. Matija Cuk of Harvard University has demonstrated how initial rotation of the Earth prior to impact could produce a disk of debris comprised of material from Earth[5]. Here we incorporate these two recent works coupled with a new gravity-centered model and outline a scenario which creates the Earth-Moon system from the collision of two planets in a single simulation. The simulation results in a large, iron-deficient Moon composed of material from both impactors. Additional results include a resultant Earth whose equatorial plane is significantly tilted off the ecliptic plane.

**The Model**
The purpose of this work was to build a model that would simulate the creation of the Earth-Moon system from a giant impact of two planets. Large aggregates of spheres with common radii, referred to as elements, were grouped together to form bodies, referred to as impactors. Following the lead of Canup, each impactor had approximately the same size and composition as present day Earth[6]. Of the elements that formed each impactor, 70 percent of the mass was from elements with a silicate material density, and 30 percent of the mass was from elements with an iron density. The total mass of each impactor was equivalent to the mass of Earth. To incorporate the work done by Cuk, the impactors were given large angular velocities[5] and then collided in various scenarios. The results of these collisions were then compared to properties of the known Earth-Moon system.

[1]Department of Engineering and Computer Science, Tarleton State University, Stephenville, Texas 76042, USA. [2]Department of Mathematics, Tarleton State University, Stephenville, Texas 76042, USA. [3]Department of Physics & Astronomy, Texas A&M University, College Station, Texas 77843, USA

Previous simulations of lunar impacts generally used smoothed particle hydrodynamics where the particles are treated as fluid elements distributed through space, and forces are calculated through pressure gradients[21]. In this work, the dynamics of the simulations were solely determined by the sum of all pair-wise forces between elements.

These pair-wise forces are described as follows. When two elements were not in contact, they interacted solely through the attractive force of gravity. Once in physical contact with each other, the elements experienced a repulsive force. The contact region of two spheres is a circle with an area determined by their separation distance. Thus, this repulsive force is proportional to the square of the separation between elements[22]. To simulate an element's resistance to deformation, each was given a shell. If this shell was not penetrated, the repulsive force was elastic. If the shell was penetrated, the force was inelastic to account for energy loss due to internal vibration, as well as deformation of the element. As two elements were forced into each other, past their shell depths, the repulsion force remained strong, but as they separated, this repulsion force was greatly reduced. This produced the inelastic component of the element-element interaction. Silicate and iron were allowed to have different repulsive strengths, different repulsion reduction amounts, and different shell depths. The force interaction between a silicate and an iron element are presented in Fig. 1, and Tables 1 and 2. In this example, it is assumed that the shell depth of iron is greater than that of silicate.

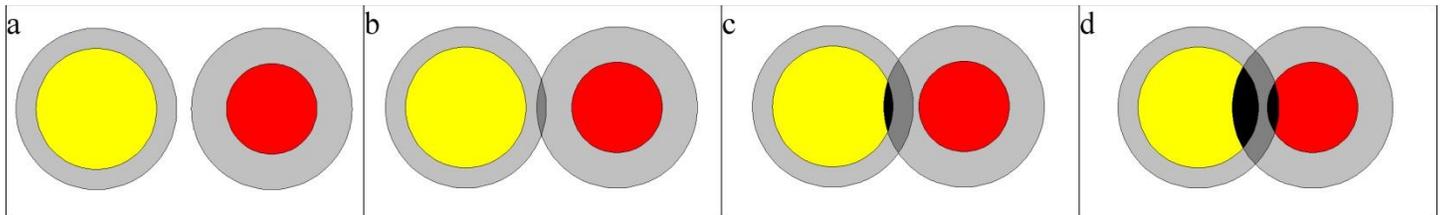

**Figure 1| Different element-element force situations**. The silicate element is yellow with a gray shell. The iron element is red with a gray shell. **a**, Elements are not in contact. **b**, Elements are in contact but neither shell is penetrated. **c**, Silicate shell has been penetrated, but not the iron shell. **d**, Both shells have been penetrated.

**Table 1| Force function Parameters**

| Parameter | Description |
|---|---|
| r | Separation between elements |
| D | Diameter of an element (iron and silicate elements have common diameters) |
| $M_{Si}$ | Mass of an silicate element |
| $M_{Fe}$ | Mass of an iron element |
| $K_{Si}$ | Repulsive parameter for silicate |
| $K_{Fe}$ | Repulsive parameter for iron-deficient |
| $KRP_{Si}$ | Percent of reduction of the silicate repulsive force |
| $KRP_{Fe}$ | Percent of reduction of the iron repulsive force |
| $SDP_{Si}$ | Shell depth percent of diameter of a silicate element |
| $SDP_{Fe}$ | Shell depth percent of diameter of an iron element |
| G | Universal gravitational constant |

**Table 2| Force function**

| Element-element separation | Force if separation is decreasing | Force if separation is increasing |
|---|---|---|
| $D \leq r$ (See Fig. 1a) | $G \cdot M_{Si} \cdot M_{Fe}/r^2$ | Same as decreasing separation |
| $D - D \cdot SDP_{Fe} \leq r < D - SDP_{Si}$ (See Fig. 1b) | $G \cdot M_{Si} \cdot M_{Fe}/r^2 - (K_{Si}+K_{Fe})(D^2 - r^2)$ | Same as decreasing separation |
| $D - D \cdot SDP_{Fe} \leq r < D - SDP_{Si}$ (See Fig. 1c) | $G \cdot M_{Si} \cdot M_{Fe}/r^2 - (K_{Si}+K_{Fe})(D^2 - r^2)$ | $G \cdot M_{Si} \cdot M_{Fe}/r^2 - (K_s \cdot KRP_{Si}+K_{Fe})(D^2 - r^2)$ |
| $epsilon \leq r < D - D \cdot SDP_{Fe}$ (See Fig. 1d) | $G \cdot M_{Si} \cdot M_{Fe}/r^2 - (K_{Si}+K_{Fe})(D^2 - r^2)$ | $G \cdot M_{Si} \cdot M_{Fe}/r^2 - (K_{Si} \cdot KRP_{Si}+K_{Fe} \cdot KRP_{Fe})(D^2 - r^2)$ |
| $r < epsilon$ | Set r equal to epsilon and calculate accordingly | Same as decreasing separation |

A check to see if the separation between two elements is less than epsilon must be performed because of the singularity in the gravitational force. If the separation between elements becomes too small, it is an indication that the repulsive parameters were not set strong enough. When this occurred it was recorded and the simulation was terminated. Forces between two silicate elements or two iron elements are similar, but only one shell depth need be considered.

To fully assess the analytic power of the model the following facts about the Earth-Moon system are given. The present day Moon orbits Earth with an eccentricity of 0.055 and an average distance from Earth of about 60 Earth radii[23]. The Moon's orbit is tidally coupled with Earth, meaning its orbital period and rotational period are equal. This is why we see only one side of the Moon. Tidal forces between Earth and the Moon created this synchronization and are still at work today adjusting the system. Tidal forces are transferring angular momentum from Earth to the Moon, causing Earth's rotation to slow and the orbital speed of the Moon to increase[5]. It is believed that the Moon accreted just outside of Earth's Roche radius which is approximately 2.9 times Earth's radius[5,17]. Moons can form only outside the Roche radius since satellites held together solely by gravity will be torn apart by Earth's tidal forces inside this region[5]. The radius of Earth is approximately 3.7 times that of the Moon. The mass of Earth is approximately 81 times that of the Moon. Earth's equatorial plane is tilted 23.4 degrees off the ecliptic plane[23].

Three examples of the model will be given. The first, a planar single-tailed collision, demonstrates how the model produced a stable Earth-Moon system from a giant impact. To date this had never been achieved in a single simulation. The second, a planar double-tailed collision, demonstrates how the model also produced a Moon composed of almost equal amounts of material from both impactors. Amazingly, the third example preserved the characteristics of the first two simulations and produces an Earth whose equatorial plane is tilted off the ecliptic plane within two degrees of the observed value.

**Planar single-tailed collision**

Most objects in our solar system reside in the ecliptic plane and orbit the sun in the same direction[3,24]. Hence, a natural scenario would be that two young rapidly spiraling proto-planets rotating and orbiting together in the ecliptic plane would have collided. The simulation depicted here is of an off-centered collision of two impactors with large opposite angular velocities, both rotating in the ecliptic plane. Because the impactors spun in opposite directions, one impactor spun into the collision while the other spun out of the collision. This caused a single-tailed spiral with the tail being composed solely of elements from the impactor spinning out of the collision (See Fig. 2).

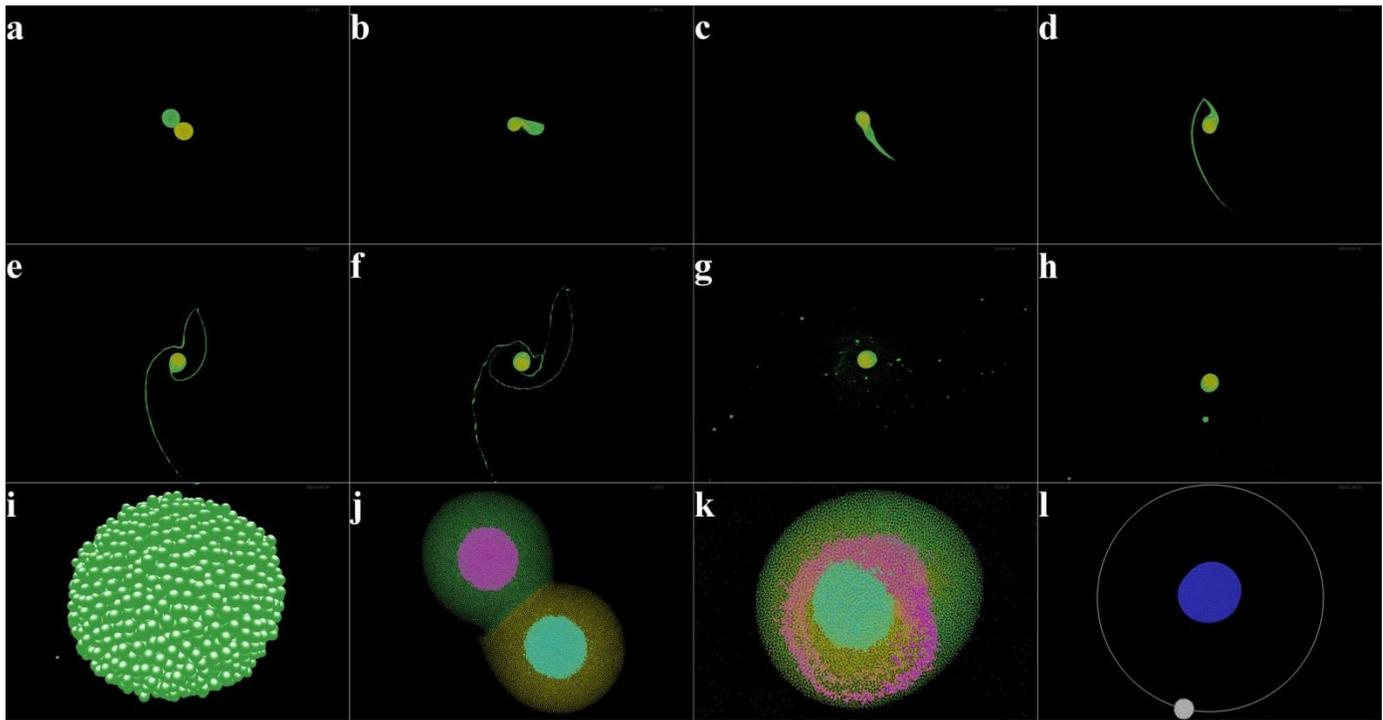

**Figure 2 | Planar single-tailed collision (131,072 elements).**
**a-h,** Top view of collision at 1.20, 2.65, 4.03, 6.48, 8.43, 10.47, 24.00, 720.00 hours, respectively, into the simulation. The green impactor is spinning clockwise. The yellow impactor is spinning counter clockwise. **i**, Close up of the resultant Moon. **j**, View of impactor cores at the point of impact. Green and yellow are silicate elements. Cyan and magenta are iron elements. **k,** View of the resultant Earth's core. **l,** Trace of resultant Moon's orbit.

The simulation ran for 720 simulated hours with the following results. A dominant Moon 1/30th the mass of the resultant Earth was formed. The Moon was void of iron. The Earth day was 4.20 hours. The orbital period of the Moon around Earth was 18.39 hours. The average distance from Earth to the Moon was 44,224.0 km. The eccentricity of the Moon's orbit was 0.078. Figure 2l shows that the Moon formed just past the Roche limit. Since the impactors were rotating in opposite directions, the total angular momentum of the system could be controlled, but the resultant Moon was composed predominantly of material from only the impactor which spun out of the collision. The initial parameters used for this simulation are provided in Supplementary Tables 1-3.

**Planar double-tailed collision**
Having completed a simulation which resulted in a stable Earth-Moon system, the focus of the next simulation was to produce a heterogeneous Moon made of material from both impactors. This would help explain the similarities of isotopes between the Earth and Moon. With the exception of Venus and Uranus, which are thought to have undergone asteroid or planetary collisions, planets in our solar system not only orbit the Sun in the same direction but also spin on their axes in the same direction[3,24]. Therefore, a more likely scenario than that causing the planar single-tailed collision is that of an off-centered collision of two young proto-planets both orbiting and spinning in the same direction in the ecliptic plane. Simulations of same-sized impactors, where both impactors spun into the collision, produced no spiral of debris and consequently no large satellites were formed. However, simulations with both impactors spinning out of the collision produced a double-tailed spiral; one spiral being composed of elements from one impactor, and the other spiral being composed of elements from the second impactor (See Fig. 3).

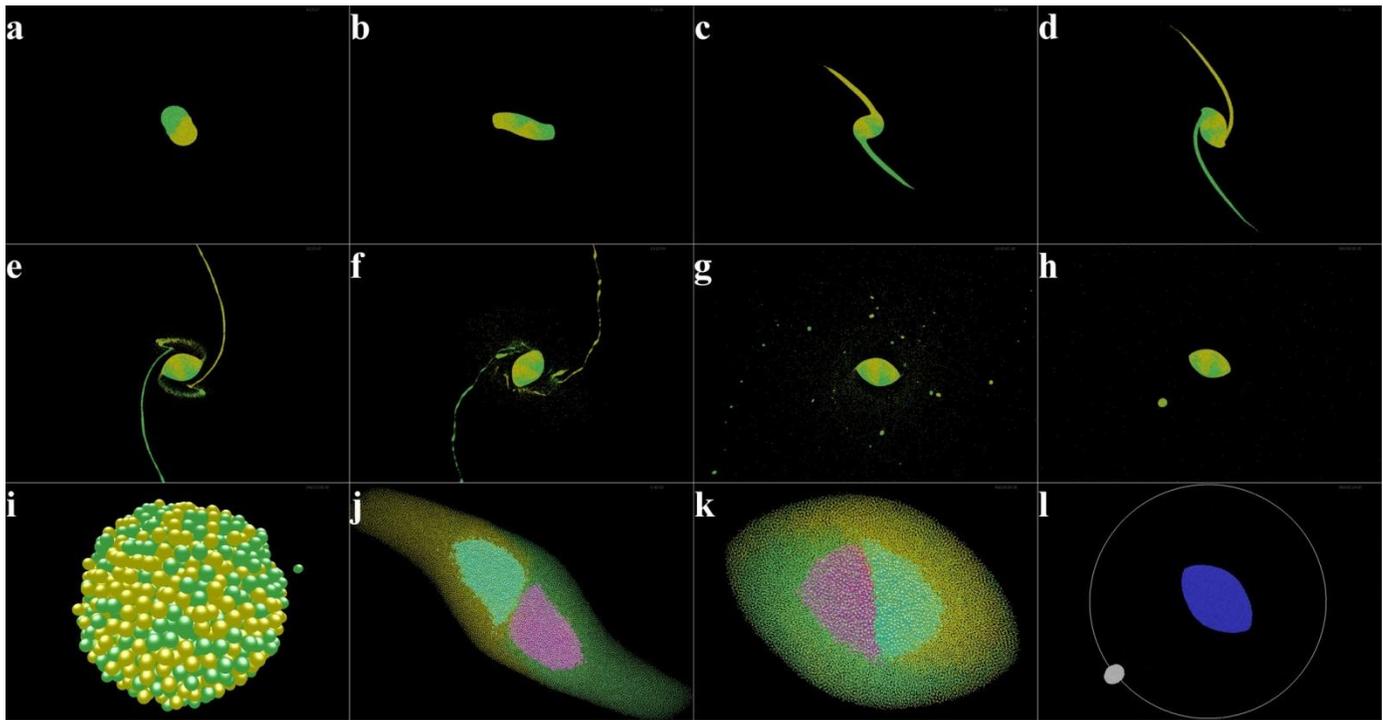

**Figure 3 | Planar double-tailed collision (131,072 elements).**
**a-h,** Top view of collision at 4.23, 5.27, 6.75, 7.88, 10.40, 13.33, 24.00, 720.00 hours respectively, into the simulation. Both impactors are spinning clockwise. **i**, Close up of the resultant Moon. **j**, View of impactor cores during impact. **k**, View of the resultant Earth's core. **l,** Trace of resultant Moon's orbit.

The simulation ran for 720 simulated hours with the following results. A dominant Moon 1/45th the mass of the resultant Earth was formed. The Moon was void of iron. The Moon was composed of almost equal amounts of material from both impactors, 48.8 percent from the yellow impactor and the remaining 51.2 percent from the green impactor. The Earth day was 3.39 hours. The orbital period of the Moon around Earth was 17.91 hours. The average distance from Earth to the Moon was 43,391.2 km. The eccentricity of the orbit was 0.077. This simulation possesses most of the qualities of the last simulation with the addition of a heterogeneous Moon. Figure 3i illustrates the uniformity of the distribution of elements from both impactors. The extreme ellipsoidal shape of the resultant Earth was a concern, but when the orbit of the Earth was slowed to a 24-hour day, Earth became spherical. The initial parameters used for this simulation are provided in Supplementary Tables 4-6.

**Parallel off-planar collision**

Having an impact scenario that produced a heterogeneous Moon, the next example demonstrates a simulation which created a resultant Earth whose equatorial plane was not in the ecliptic plane. Though it is likely that two proto-planets would be orbiting and rotating in the same direction, parallel to the ecliptic plane, it is not likely that they would both be rotating exactly in the ecliptic plane. Therefore, a more likely scenario than that depicted in the planar double-tailed collision would be that of two proto-planets, both rotating and orbiting in the same direction, parallel to the ecliptic plane, but not co-planar. If two such impactors collided in an off-centered collision with both impactors spinning out of the collision, a double-tailed spiral was again formed, but the resultant Earth's equatorial plane was tilted off the ecliptic plane (See Fig. 4).

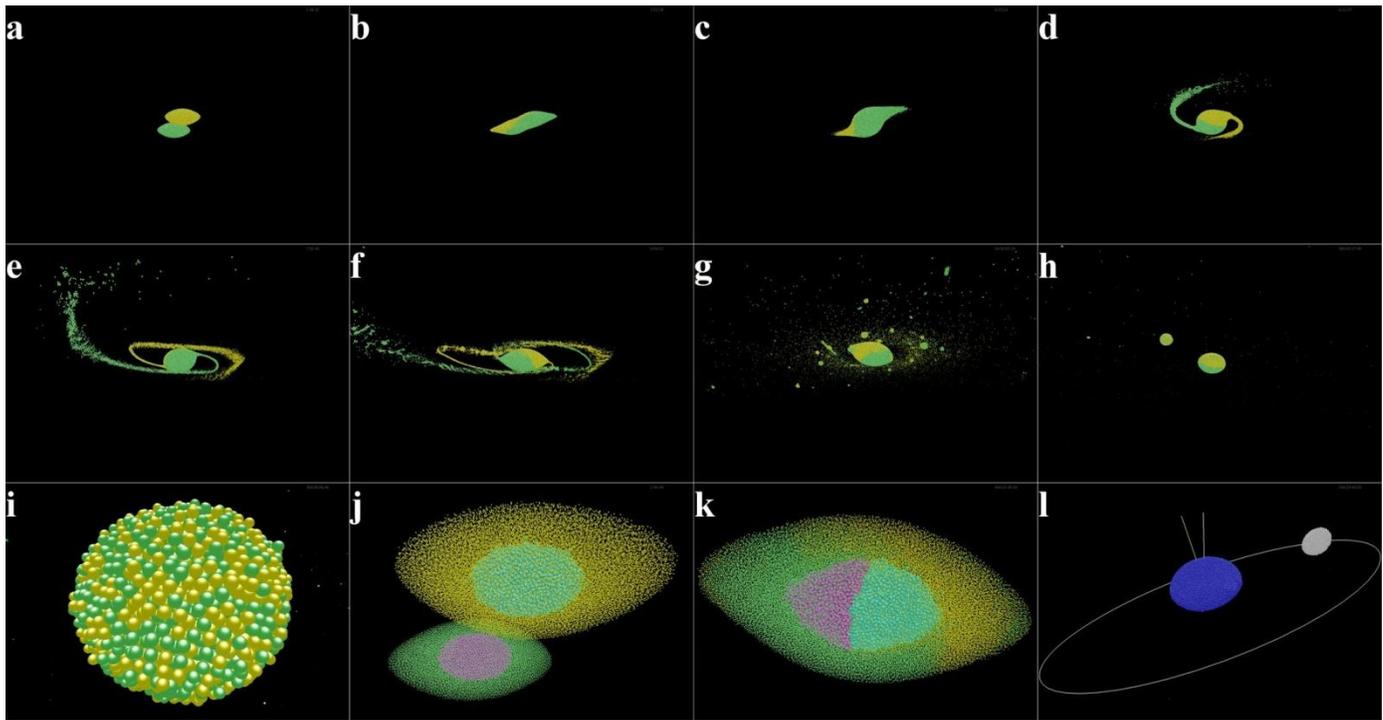

**Figure 4 | Parallel off-planar collision (131,072 elements).**
**a-h,** Side view of collision at 2.82, 3.87, 4.42, 6.17, 7.88, 9.07, 24.00, 720.00 hours respectively, into the simulation. Both impactors would be spinning clockwise if viewed as in Figure 3. **i,** Close up of the resultant Moon. **j,** View of impactor cores at impact. **k,** View of the resultant Earth's core. **l,** Trace of resultant Moon's orbit and tilt off ecliptic. The white ray is normal to the ecliptic plane, and the yellow ray is normal to the Earth's equatorial plane.

The simulation ran for 720 simulated hours with the following results. A dominant Moon 1/26th the mass of the resultant Earth was formed. The Moon was void of iron. The Moon was composed of almost equal amounts of material from both impactors, 48.7 percent from the yellow impactor and the remaining 51.3 percent from the green impactor. The Earth day was 3.39 hours. The orbital period of the Moon around Earth was 13.88 hours. The average distance from Earth to the Moon was 36,672.9 km. The eccentricity of the orbit was 0.144. The resultant Earth's equatorial plane was 21.51 degrees off the ecliptic plane, only 1.89 degrees off the known value. The initial parameters used for this simulation are provided in Supplementary Tables 7-9.

A phenomenon that is worth noting in each of the simulations is the kink in the spiral tails. This occurs as gravity pulls the core elements out of the base of the spiral tails. It was observed that when this kinking occurred, a stable Earth-Moon system was more likely to evolve. The kink is most pronounced in Figure 2d.

**Discussion**
This work adds even more credibility to the giant impact hypothesis proposed almost 40 years ago. In addition to producing a large, iron-deficient Moon from a giant impact, these examples show how an impact could create a heterogeneous Moon made of material from both impactors and a resultant Earth whose equatorial plane is tilted substantially off the ecliptic plane.

Simulations were run with a variety of initial conditions: force repulsion parameters, numerical techniques, number of elements used, densities of different types of silicate material, densities for different states of iron, and time step sizes. A large number of these produced excellent Earth-Moon systems, some more accurate than the ones presented in one or more aspects. However, the presented simulations demonstrated particular aspects that were of interest for this research, such as the heterogeneous structure of the Moon and the equatorial tilt off the ecliptic plane.

Long run simulations are presently being generated to study tidal coupling and the transfer of angular momentum from Earth to the Moon. Work is also being done to quantify the repulsive parameters of the model. It is our anticipation that the computational simplicity of the model and its ability to capture so much observed phenomena will attract more researchers to the exciting field of computational astrophysics.

**METHOD SUMMARY**

Gravity was not approximated in the model; hence the computational demand grew at a rate of the number of elements squared. The amount of computational power required to perform this work would have been out of the financial scope of this project without the recent advancement in general purpose graphic processing units (GPGPUs)[25]. All simulations were run on a single NVIDIA 580 or 680 graphics card. The numerical algorithms were written in C, C++, and Compute Unified Device Architecture (CUDA)[25]. The graphical algorithms were written in C, C++ and OpenGL. The numerical schemes used were the Leap-frog formulas[26], and Kutta's fourth-order formula[27]. The exploratory work was done with 16,384 elements and a large time step to search for promising force parameters and initial conditions. These were then used in larger simulations with smaller time steps for increased accuracy and proof of scalability. The power of the NVIDIA graphics cards allowed the exploratory work to be done in real time which was instrumental in successfully implementing the model.

**Supplementary Information** is available in the online version of this paper.

**Acknowledgements** We would like to thank NVIDIA and Mellanox Technologies for their donation of hardware to support the project.

**Author Contributions** J.C.E., T.C.S., B.H.H., and J.L.H. conducted research. B.H.H. and J.L.H. secured funding. J.C.E., T.C.S., B.H.H. and J.L.H. wrote the paper. B.M.W. supervised project.

**Author Information** Reprints and permissions information is available at www.nature.com/reprints. The authors declare no competing financial interests. Readers are welcome to comment on the online version of the paper. Correspondence and requests for materials should be addressed to B.M.W. (wyatt@tarleton.edu).

**Supplementary Table 1| Parameters for Planar single-tailed collision**

| Parameter | Value |
| --- | --- |
| D | 376.78 km |
| $M_{Si}$ | $7.4161 \cdot 10^{19}$ kg |
| $M_{Fe}$ | $1.9549 \cdot 10^{20}$ kg |
| $K_{Si}$ | $2.9114 \cdot 10^{11}$ kg/(m s$^2$) |
| $K_{Fe}$ | $5.8228 \cdot 10^{11}$ kg/(m s$^2$) |
| $KRP_{Si}$ | 0.01 |
| $KRP_{Fe}$ | 0.02 |
| $SDP_{Si}$ | 0.001 |
| $SDP_{Fe}$ | 0.002 |
| epsilon | 47.0975 km |
| Time step | 5.8117 s |
| Numerical technique | Leap-Frog formulas |

**Supplementary Table 2| Initial values for Planar single-tailed collision (yellow impactor)**

| Parameter | x value | y value | z value |
| --- | --- | --- | --- |
| Center of mass | 23925.0 km | 0.0 km | 9042.7 km |
| Linear velocity | -3.2416 km/s | 0.0 km/s | 0.0 km/s |
| Angular velocity | 0.0 rad/h | 3.0973 rad/h | 0.0 rad/h |

**Supplementary Table 3| Initial values for Planar single-tailed collision (green impactor)**

| Parameter | x value | y value | z value |
| --- | --- | --- | --- |
| Center of mass | -23925.0 km | 0.0 km | -9042.7 km |
| Linear velocity | 3.2416 km/s | 0.0 km/s | 0.0 km/s |
| Angular velocity | 0.0 rad/h | -3.0973 rad/h | 0.0 rad/h |

**Supplementary Table 4| Parameters for Planar double-tailed collision**

| Parameter | Value |
|---|---|
| D | 376.78 km |
| $M_{Si}$ | $7.4161 \cdot 10^{19}$ kg |
| $M_{Fe}$ | $1.9549 \cdot 10^{20}$ kg |
| $K_{Si}$ | $7.2785 \cdot 10^{10}$ kg/(m s$^2$) |
| $K_{Fe}$ | $2.9114 \cdot 10^{11}$ kg/(m s$^2$) |
| $KRP_{Si}$ | 0.01 |
| $KRP_{Fe}$ | 0.02 |
| $SDP_{Si}$ | 0.001 |
| $SDP_{Fe}$ | 0.01 |
| epsilon | 47.0975 km |
| Time step | 5.8117 s |
| Numerical technique | Leap-Frog formulas |

**Supplementary Table 5| Initial values for Planar double-tailed collision (yellow impactor)**

| Parameter | x value | y value | z value |
|---|---|---|---|
| Center of mass | 37678.0 km | 0.0 km | 9042.7 km |
| Linear velocity | -1.29678 km/s | 0.0 km/s | 0.0 km/s |
| Angular velocity | 0.0 rad/h | 3.0973 rad/h | 0.0 rad/h |

**Supplementary Table 6| Initial values for Planar double-tailed collision (green impactor)**

| Parameter | x value | y value | z value |
|---|---|---|---|
| Center of mass | -37678.0 km | 0.0 km | -9042.7 km |
| Linear velocity | 1.29678 km/s | 0.0 km/s | 0.0 km/s |
| Angular velocity | 0.0 rad/h | 3.0973 rad/h | 0.0 rad/h |

**Supplementary Table 7| Parameters for Parallel off-planar collision**

| Parameter | Value |
|---|---|
| D | 376.78 km |
| $M_{Si}$ | $7.4161 \cdot 10^{19}$ kg |
| $M_{Fe}$ | $1.9549 \cdot 10^{20}$ kg |
| $K_{Si}$ | $7.2785 \cdot 10^{10}$ kg/(m s$^2$) |
| $K_{Fe}$ | $2.9114 \cdot 10^{11}$ kg/(m s$^2$) |
| $KRP_{Si}$ | 0.01 |
| $KRP_{Fe}$ | 0.02 |
| $SDP_{Si}$ | 0.001 |
| $SDP_{Fe}$ | 0.01 |
| epsilon | 47.0975 km |
| Time step | 5.8117 s |
| Numerical technique | Leap-Frog formulas |

**Supplementary Table 8| Initial values for Parallel off-planar collision (yellow impactor)**

| Parameter | x value | y value | z value |
|---|---|---|---|
| Center of mass | 24490.7 km | 9042.7 km | 9042.7 km |
| Linear velocity | -1.29664 km/s | 0.0 km/s | 0.0 km/s |
| Angular velocity | 0.0 rad/h | 3.0407 rad/h | 0.0 rad/h |

**Supplementary Table 9| Initial values for Parallel off-planar collision (green impactor)**

| Parameter | x value | y value | z value |
|---|---|---|---|
| Center of mass | -24490.7 km | -9042.7 km | -9042.7 km |
| Linear velocity | 1.29664 km/s | 0.0 km/s | 0.0 km/s |
| Angular velocity | 0.0 rad/h | 3.0407 rad/h | 0.0 rad/h |